\documentclass{llncs}

\usepackage{makeidx}  
\usepackage{url,color,algorithmic,algorithm}
\usepackage{graphicx}

\usepackage{subfigure}

\begin{document}

\frontmatter          

\pagestyle{headings}  
\addtocmark{Cognitive challenges} 

\title{On the role of working memory in trading-off skills and situation awareness in Sudoku}


\titlerunning{Working memory, skills and situation awareness in Sudoku}

\author{George Leu \and Jiangjun Tang \and Hussein Abbass}

\authorrunning{Leu, Tang and Abbass} 

\tocauthor{George Leu, Jiangjun Tang, and Hussein Abbass}

\institute{School of Engineering and Information Technology\\
University of New South Wales, ADFA, Canberra, Australia\\
\email{(G.Leu,J.Tang,H.Abbass)@adfa.edu.au}}

\maketitle              

\begin{abstract}
\noindent Working memory accounts for the ability of humans to
perform cognitive processing, by handling both the representation
of information (the mental picture forming the situation awareness)
and the space required for processing these information (skill processing). The more complex the skills are, the more processing space they require, the less space becomes available for storage of information. This interplay
between situation awareness and skills is critical in many applications. Theoretically, it is less understood in cognition and neuroscience. In the meantime, and practically, it is vital when analysing the mental processes involved in safety-critical domains.

\noindent In this paper, we use the Sudoku game as a vehicle to
study this trade-off. This game combines two features that are
present during a user interaction with a software in many safety
critical domains: scanning for information and processing of
information. We use a society of agents for investigating how this
trade-off influences player's proficiency.

\keywords{cognitive challenge, scanning skills, working memory}
\end{abstract}

\section{Introduction}

\noindent We investigate the trade-off between the processing and
storage functions of working memory in Sudoku, one of the most
popular puzzles. During a game, players visually scan
the Sudoku grid in a continuous manner searching for cells
containing information that can be propagated throughout the grid
to narrow down the degrees of freedom of empty cells. These
scanning skills are an integral part of the skill set required for
solving the Sudoku puzzle, and the players learn gradually how to
use them properly.

\noindent Since our interest is to understand the cognitive
functions and working memory within Sudoku, we only consider
players playing the game without note-taking abilities. External
note-taking extends players' working memory with an external
holder of information. This type of game play is not suitable for
this investigation because it does not give us the right insight
into the player's cognitive capacity.

\noindent Depending on the complexity of the scanning patterns
used for improving situation awareness, the scanning activity can
become strenuous. This is because the player must use
the working memory for handling both the current scanning task
(space for executing skills), and the information resulted from
scanning (situation awareness). Between the two functions of the
working memory, there is always a trade-off \cite{Baddeley2001},
given its finite capacity. Understanding how these two functions interact would help us in designing training and education programs that are cognitively plausible and that account for this trade-off.

\noindent The rest of this paper is structured as follows:
Section~\ref{01Background} presents background information on
working memory, followed by the methodology in
Section~\ref{02Methodology}, results in Section~\ref{03Results},
then conclusions are drawn in Section~\ref{04Conclusion}.

\section{Background on working memory}\label{01Background}

\noindent Early memory models emphasized one of the two
functions of working memory. On one side, the multi-store model \cite{Atkinson1968}, suggested that memory was a series of stores, i.e. the sensory memory, the short-term memory, and the long term memory, focusing on the storage function of memory. On the other side, the ``levels of processing" model \cite{Craik1972} concentrates on the processes involved in memory. It considers memory as a consequence of the depth of information processing, with no clear distinction between short-term memory, long-term memory or other stores.

\noindent Later, Baddeley et al. \cite{Baddeley1974} introduced
the concept of Working Memory (WM) and showed that short-term
memory is more than just one simple store. WM is still
short-term memory, but instead of all information going into one
single store, there are different subsystems with different
functions (e.g. visual, auditory, etc.). The
resource-sharing approach on memory adds on Baddeley's work and
emphasizes the ``storage versus processing" paradigm. Case et al.
\cite{Case1982} consider that WM accounts for the
processing resources of an individual, and is the sum of a storage
space and an operating space. They show that the more complex the
current processing task is, the more operating space it uses,
leaving less space for information storage. This was
further supported by other studies which assumed that human
performance in various cognitive tasks is strongly related to
working memory \cite{Lovett1999,Baddeley2001}.

\noindent Another aspect involved in the study of working memory
is the forgetting mechanism associated with its normal operation.
The displacement theory \cite{Atkinson1968} assumes the working
memory is a first-in-first-out (FIFO) queue with limited capacity,
and explains how the most recent information stored in the memory
is easier to be recalled (recency effect). The trace decay theory
\cite{Towse1995} assumes that the events between learning and
recall have no effect on recall, the essential influencing factor
being the period of time the information has been retained. The
interference theory challenges the decay concepts, and attributes
forgetting to the existence of interference
\cite{Chandler1989,Brown2007}. A review of the forgetting theories
can be found in \cite{Wixted2004}.

\noindent From a computational perspective, numerous studies
proposed various combinations of working memory models and
theories of forgetting, in order to instantiate plausible
behaviors. Two major approaches have been proposed over the years:
the localist and the distributed representation of working memory.
In the localist approach, the memory is seen from an item-unit
perspective, in which the informational items are stored in
corresponding units situated in certain locations of the memory.
The Competitive Queueing \cite{Burgess2005} and the Primacy
\cite{Page1998} models propose time-related displacement and
displacement only, respectively, whereas the Start-End
\cite{Henson1998} model proposes a multi-level displacement-based
approach. The distributed approach sees an informational item as
distributed throughout the memory space rather than as in the
item-unit equality view. Thus, in the distributed approaches, the
focus shifts from the localist practice of retrieving items from
their corresponding locations, to identifying patterns of
activation which recompose informational items from their parts
distributed through multiple layers of interconnected units. A
displacement-based distributed approach was proposed by
Lewandowsky and colleagues \cite{Lewandowsky1999} based on the
Theory of Distributed Associative Memory. Later, models such as
OSCAR \cite{Brown2000} and SIMPLE \cite{Brown2007} considered
interference-based implementations in which the working memory is
hierarchical in structure, oscillatory in time, and contextually
activated.

\section{Methodology}\label{02Methodology}

\noindent From the Sudoku puzzle perspective the scanning skills
are forming situation awareness picture required to solve the
Sudoku game, which is further stored in the WM. The
scanning pattern is based on Sudoku cells belonging to the grid.
This suggests a localist approach for the WM, in which each step in the scanning pattern produces a scanned item corresponding to a memory unit. We consider that the items are stored in a FIFO queue, as in the displacement theory.

\begin{figure}[!hpbt]
    \center
    \subfigure[The learning framework]{
        \includegraphics[width=0.4\textwidth]{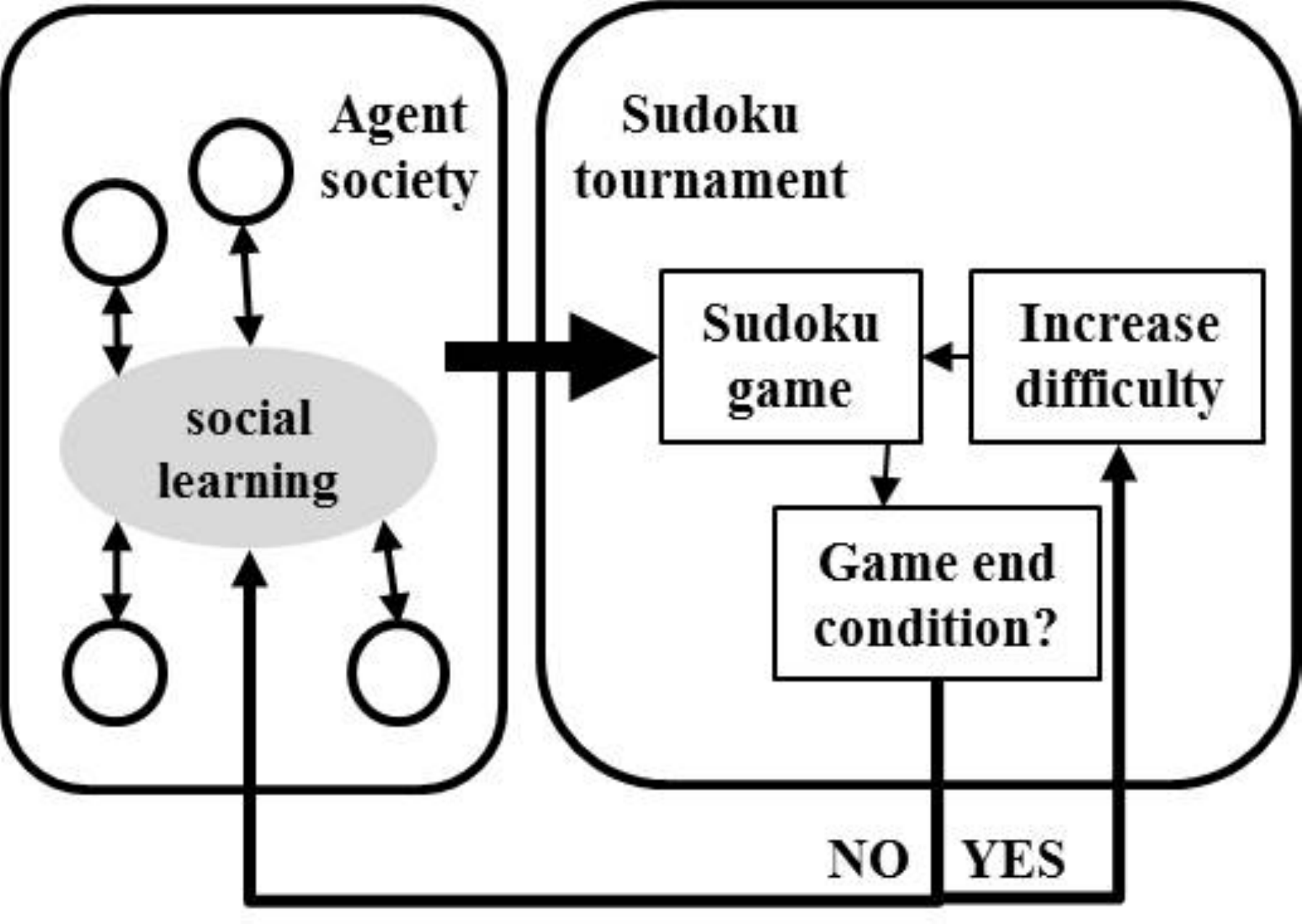}
        \label{fig_Framework}
    }
    ~
    \subfigure[The agent]{
        \includegraphics[width=0.4\textwidth]{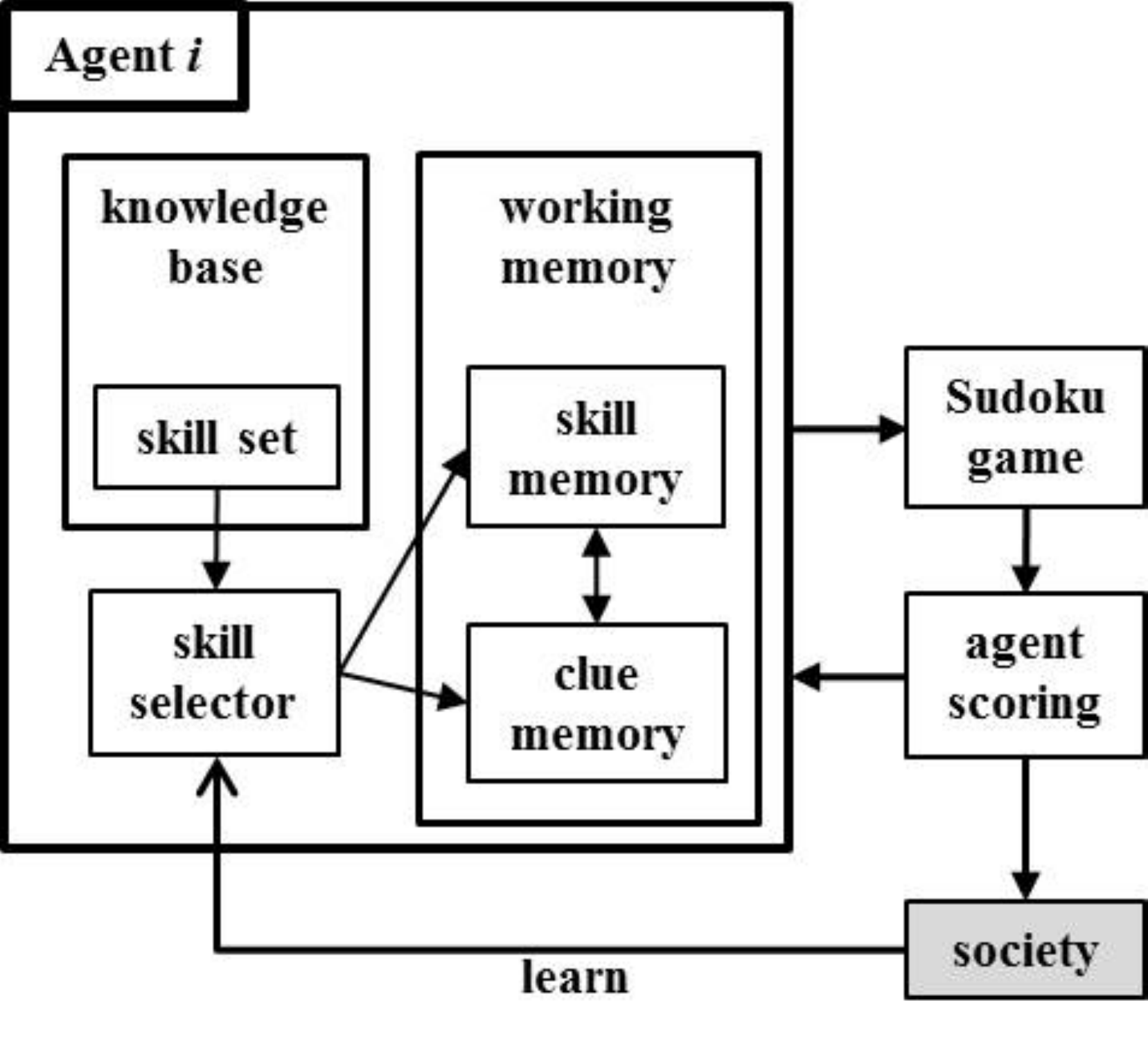}
        \label{fig_Agent}
    }
    \caption{Methodological approach.}\label{fig_Method}
\end{figure}

\noindent Figure ~\ref{fig_Method} presents the social learning
framework and the internal structure of an agent. The agents in
the society are endowed with fixed working memory and adaptive
ability to choose the scanning skills to be loaded into memory and
get executed. Their resultant Sudoku proficiency is tested in a
tournament with several rounds of increasing difficulty. In each
round one game is proposed, which all agents try to solve. If no
agent is able to successfully fill the grid at the end of the game
in the first attempt, the agents learn socially from each-other
how to adapt their skill selection towards a better proficiency,
and the game is replayed in the new conditions. The learning
process continues until at least one agent becomes proficient
enough to complete the grid or until learning does not bring any
more improvement. The tournament continues with the next round
where a game with increased difficulty is proposed and the process
is repeated.

\subsection{The agent}

\noindent {\bf The scanning skill set: } Each agent can choose
from a set of scanning skills, constant over the whole society.
The complexity of a skill can be related to the amount of
information that must be stored in the working memory to describe
the scanning pattern. The skill set shown below presents the number of scanned cells and the number of memory units needed for storing them. Since we adopt a localist approach on working memory, the two numbers are equal. Each skill is named as COL for scanning a column, ROW for scanning a row, or BOX for scanning a box. The subsequent digit in the name represents the number of dimension of this scanning activity; thus, the size of the storage required to perform this scanning task.
{\center{
        \begin{tabular}{|l|c|c|c|c|c|c|c|c|c|c|}\hline
            Skill code & 1 & 2 & 3 & 4 & 5 & 6 & 7 & 8 & 9 & 10\\ \hline
            Skill name & ROW3 & ROW5 & ROW7 & ROW9 & COL3 & COL5 & COL7 & COL9 & BOX5 & BOX9\\ \hline
            Grid cells & 3 & 5 & 7 & 9 & 3 & 5 & 7 & 9 & 5 & 9\\ \hline
            Mem. units & 3 & 5 & 7 & 9 & 3 & 5 & 7 & 9 & 5 & 9\\ \hline
        \end{tabular}
        }}

\noindent The skill selector is a vector $V=V(v_1,\dots,v_{10})$
containing the weights associated with each skill in the skill
set, where $v_i\in[0,1]$. The agent loads the first $m$ skills
with the highest weight that fit in the skill memory $M_s$. The
selection vector is initialized at the beginning of the simulation
for each agent, then it is updated during the tournament as part
of the learning process.

\noindent {\bf The working memory: } The working memory has two
components: the skill memory $(M_s)$, and the situation awareness
memory $(M_c)$. The sizes of the two memory components are
predefined for each agent throughout the tournament, and they
differ from one agent to another, but their sum is identical for
all agents in the society, as shown in equation \ref{eq_memory}.
In other words, the total working memory space is maintained
constant for each agent in the society.

\begin{equation}
    M=M_i=M_{s_i}+M_{c_i}, \forall{i=1,n}; \label{eq_memory}
\end{equation}

\noindent The skill memory stores the representation of the skills
selected by the skill selector. At the end of each game or round,
the skill memory is erased and then reloaded with the new skills
selected by the skill selector as a result of the learning
process. The new skills are to be used in the next game if the
game must be replayed or in the next round if a new round must
start. Thus, the skill memory is rewritten through incremental
social learning, each time a game is replayed in a round of the
tournament.

\noindent The situation awareness memory stores the results of the
scanning process and is modelled from a displacement theory
perspective, as a FIFO queue of size $M_c$. In this queue, the
information discovered by applying the skills currently stored in
skill memory are stored in continuation of those stored at
previous step. Thus, older information, which exceeds the queue
size is pushed out and lost.

\subsection{The tournament}

\noindent {\bf Playing a game: } During a game an agent scans the
Sudoku grid in order to form its situation awareness picture which
allows the propagation of Sudoku constraints/rules. The agent
visits each empty cell of the grid and applies to it the selected
skills. After the agent applies the skills on all empty cells and
propagates the Sudoku rules, a score is given to the game
reflecting on agent's proficiency.

\noindent {\bf Scoring a game: } Agents receive scores based on
the remaining degrees of freedom of empty cells in the grid after
a fixed number of allowed steps. The degree of freedom for an
empty cell, $f_c$, is the number of possible candidates found
after propagation of domains. If the Sudoku grid is complete at
the end of the game, there is no degree of freedom left. If the
grid still has empty cells, the degrees of freedom in each empty
cell are added, generating $f_i=\sum{f_c}$, the total degree of
freedom for agent $i$. The performance of an agent is inversely
proportional to $f_i$: the less degrees of freedom remaining at
the end of a game, the better the agent performs. Thus, the score
$s_i$ is defined in Equation~\ref{eq_score}, where $f_i$ is the
total degrees of freedom for agent $i$, and $f_{max}$ is the
maximum total degrees of freedom over the society.

\begin{equation}
    s_i=1-\frac{f_i}{f_{max}}; \label{eq_score}
\end{equation}

\noindent {\bf Learning: } At the end of a game, with or without
completion of the grid, each agent updates its skill selection
vector using the experience of the other agents. The update can be
viewed from two perspectives, which we combine: the current agent
searches for the agent with the highest score and most similar
memory. Thus, the agents in the society learn from other agents
with as similar memory (cognitive capacity) as possible, and score
as high as possible.

\noindent First, we define a similarity metric $D_{ij}$ for
working memory of agents, as in Equation~\ref{eq_memDist}, where
$M_{c_i}$ and $M_{c_j}$ are the awareness memory size of agents $i$
(current agent) and $j$ (an agent from the society) respectively,
and $M_{c_{max}}$ and $M_{c_{min}}$ are the maximum and minimum
sizes of the awareness memory. If $i=j$ then the memory similarity of
an agent to itself $D_{ii} = 1$.

\begin{equation}
    D_{ij}=1-\vert{\frac{M_{c_i}-M_{c_j}}{M_{c_{max}}-M_{c_{min}}}}\vert; \label{eq_memDist}
\end{equation}

\noindent Then, we couple the memory similarity metric with the
score, as in Equation \ref{eq_fitness}, where $F_{ij}$ is the
fitness between current agent $i$ and an agent $j$ from the
society. The agent $j$ corresponding to the $\max{F_{ij}}$ will be
the agent in the society which is best fit for participating in
agent $i$ learning process. We note that the fitness $F_{ii}$ of
an agent to itself equals its score in the Sudoku game, since
$D_{ii}=1$.

\begin{equation}
    F_{ij} = s_j \times D_{ij}; \label{eq_fitness}
\end{equation}

\noindent The amount of participation of agent $j$ in agent $i$'s
learning is given by the actual value of the maximum fitness
coefficient, hence, the update function becomes as in Equation
\ref{eq_final_update} and applies when $F_{ij}>F_{ii}$.

\begin{equation}
    v_i(t+1)=(1-F_{ij})v_i(t)+F_{ij}v_j(t); \label{eq_final_update}
\end{equation}

\subsection{The experimental setup}

\noindent The skill selection vector is initialized with
random weights for each agent at the beginning of tournament. Ten sets of experiments are run with different
seeds.

\noindent The total working memory $M$ for each agent in the
society is 54 units, corresponding to 54 digits associated with
the Sudoku puzzle. Within the total memory, the skill memory $M_s$
and the situation awareness memory $M_c$ can be tuned in the range
between 9 and 45. The agents in the society are endowed with a
situation awareness memory ranging between 9 to 45 progressively
with an increment of 2. The skill memory follows the
opposite variation pattern. Consequently, the society has 19
agents.

\noindent The tournament consists of nine rounds of increasing
level of difficulty, where difficulty is associated with the
initial number of non-empty cells existent in the grid. In this
study, grids with 76, 74, 71, 67, 62, 56, 49, 41 and 32 initial
non-empty cells are used for the 9 tournament rounds in that
order.

\section{Results and discussion}\label{03Results}

\begin{figure}[t]
    \center
    \subfigure[The popular skills over the tournament]
    {
        \includegraphics[width=0.45\textwidth,height=0.18\textheight]{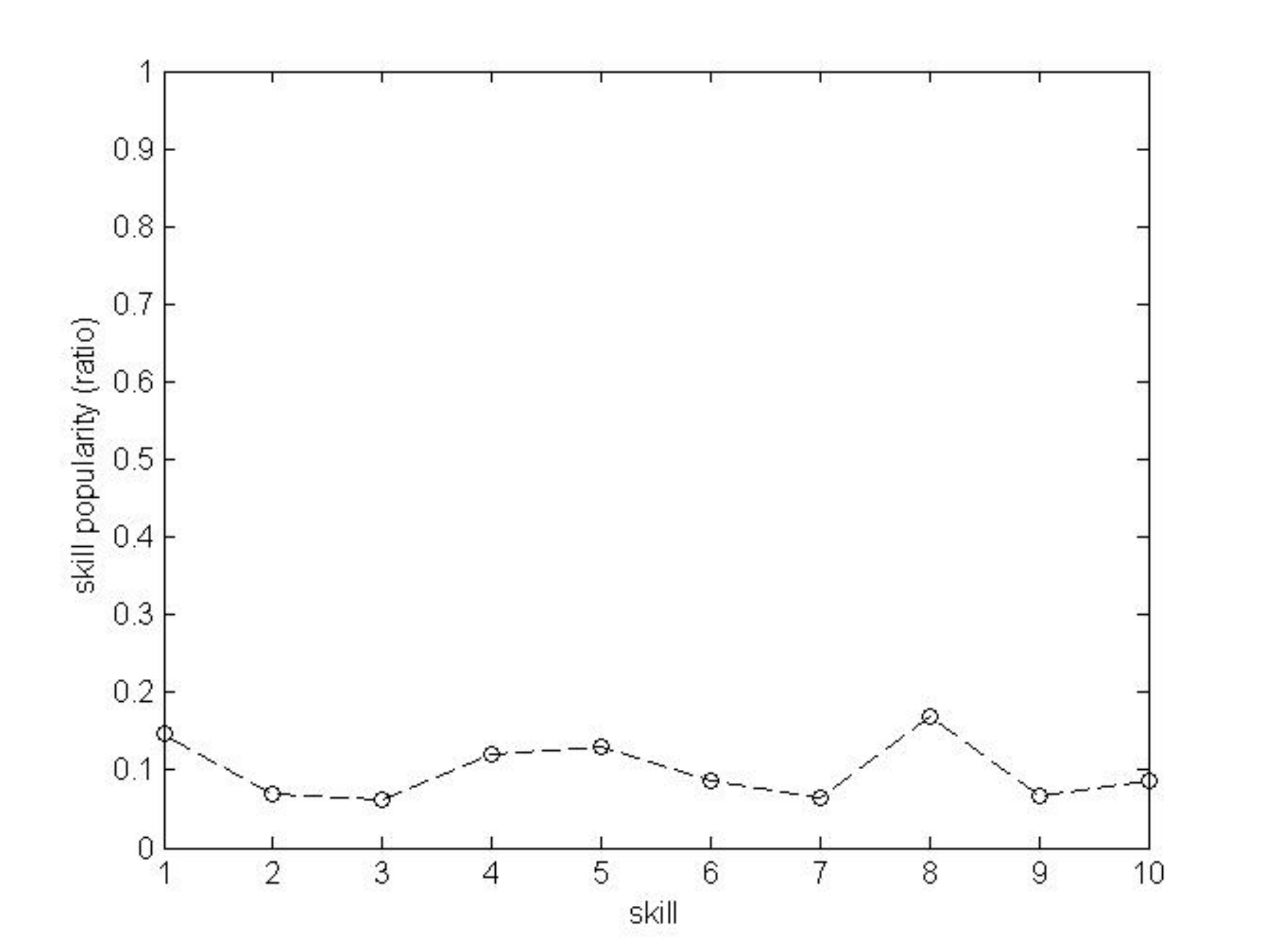}
        \label{fig_SkillPop}
    }
    ~
    \subfigure[The effective skills over the tournament]
    {
        \includegraphics[width=0.45\textwidth,height=0.18\textheight]{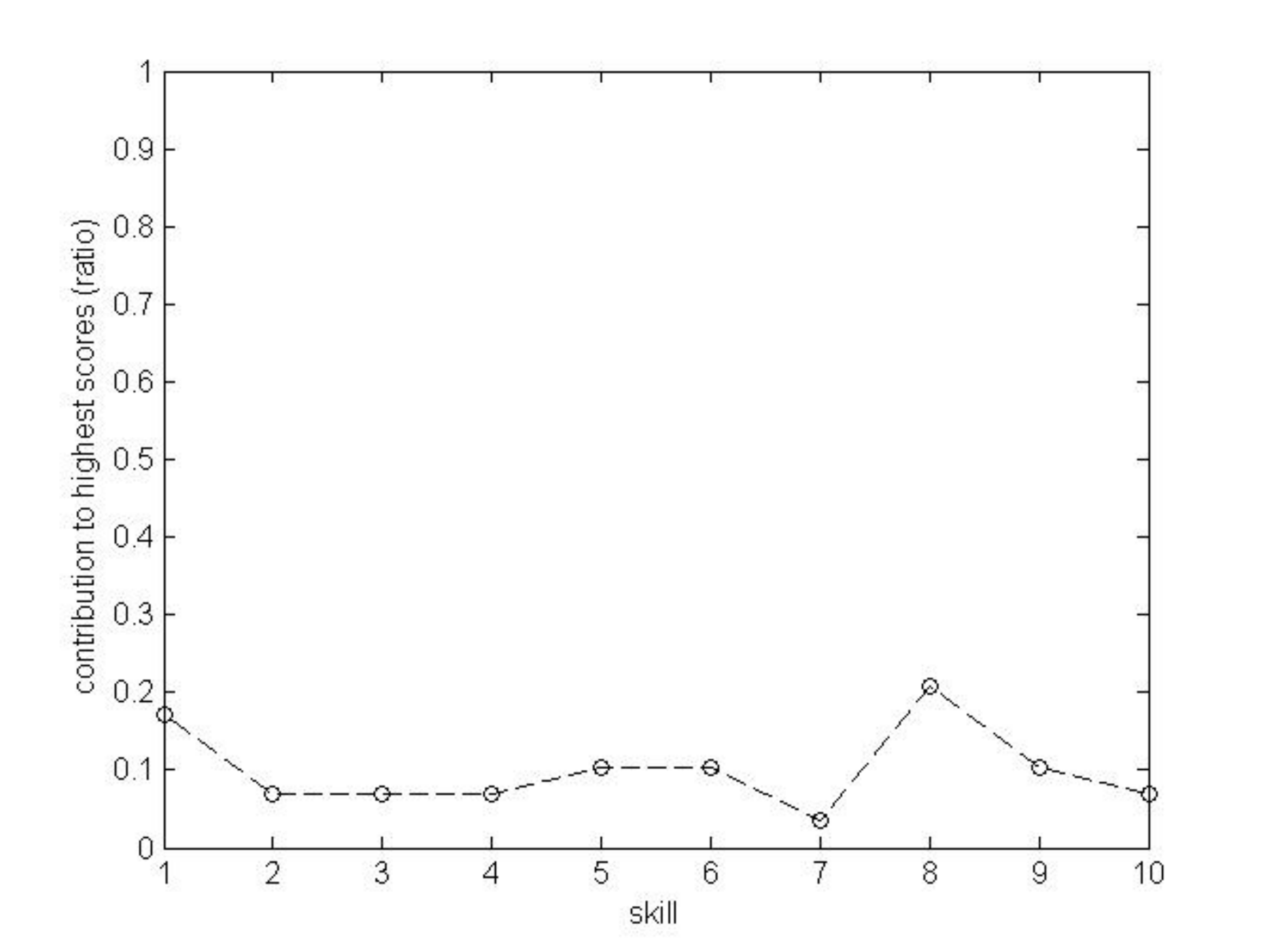}
        \label{fig_SkillContr}
    }
    \subfigure[Usage of most effective skill over all memory setups]
    {
        \includegraphics[width=0.45\textwidth,height=0.18\textheight]{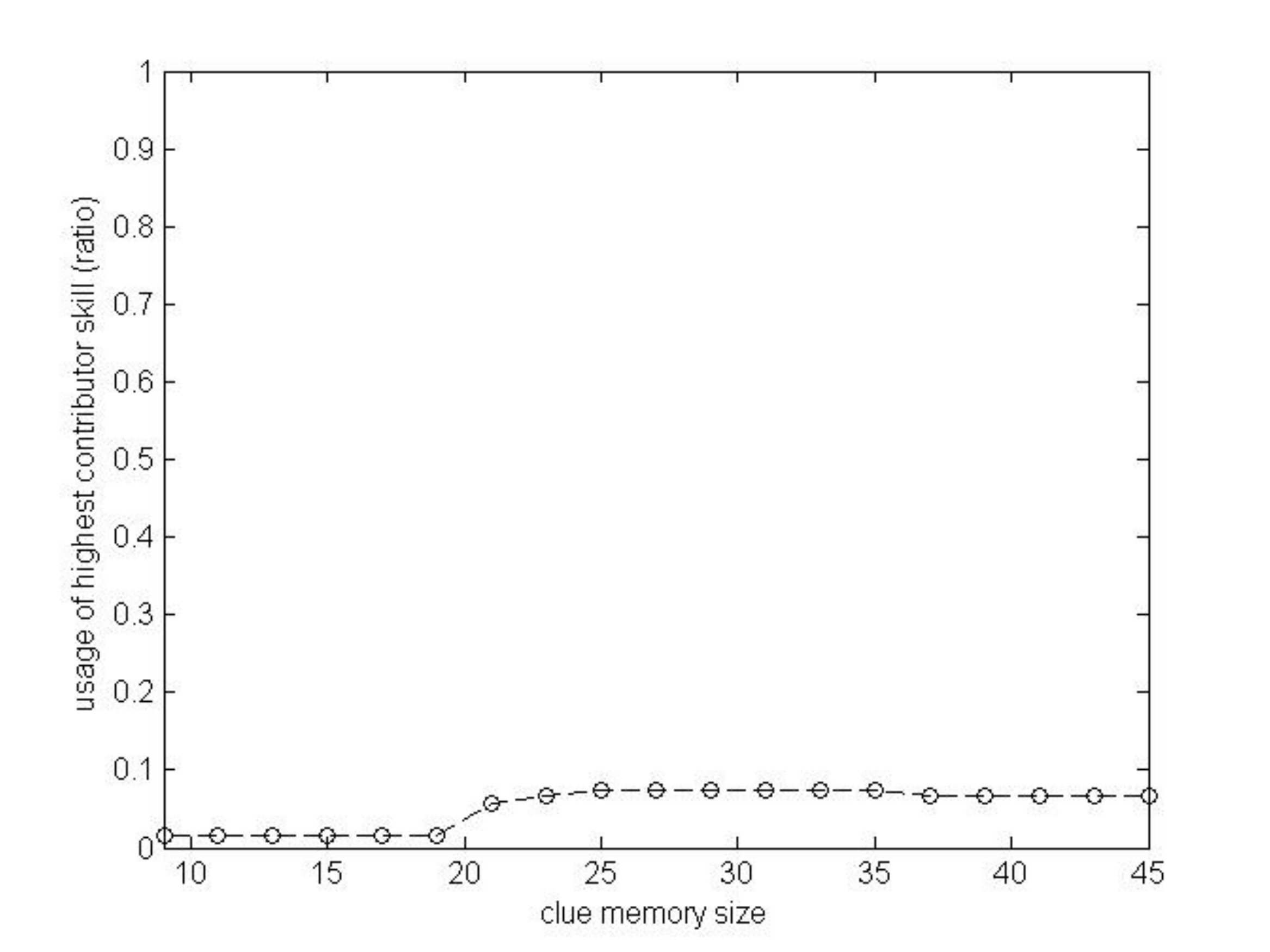}
        \label{fig_SkillMaxMem}
    }
    ~
    \subfigure[Performance throughout the tournament for all memory setups]
    {
        \includegraphics[width=0.45\textwidth,height=0.18\textheight]{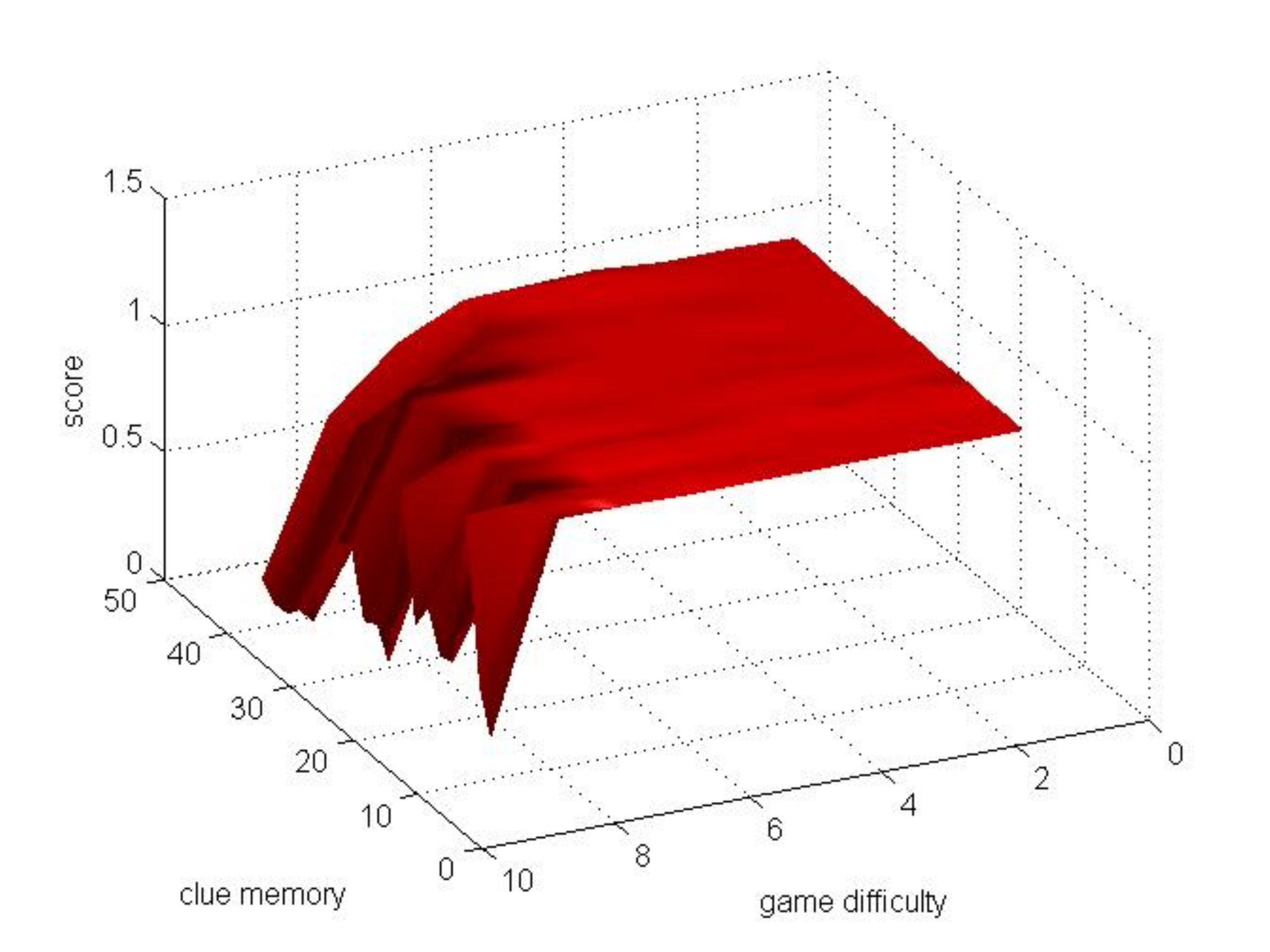}
        \label{fig_MemGameScore}
    }
    \caption{Skills over the tournament.}\label{fig_Results1}
\end{figure}

\noindent Figure~\ref{fig_SkillPop}, shows that skill 8 (COL9) is
the most used by the agent society throughout the Sudoku
tournament, followed by skill 1 (ROW5). However, a skill being
chosen may not imply it produces high scores in Sudoku games. We
test this by investigating what are the skills most used for
obtaining the highest scores. Figure~\ref{fig_SkillContr}
demonstrates that indeed skill 8 and skill 1 are also the most
effective scanning skills, given the experimental setup used in
this paper. We understand that skill 8 is the most effective
skill, followed by skill 1.

\noindent Further, we select the skill with highest proficiency
(skill 8) and investigate for which size of the $M_c$ it is most used. This is equivalent to searching which agent used this skill most, since the agents differentiate from each other through the ratio between skill and situation awareness memory. Figure~\ref{fig_SkillMaxMem} shows that the
highest effective scanning skill is used more by agents with high amount of memory reserved for situation awareness storage, and less used by other agents.


\noindent We continue to investigate the overall influence on
performance of the scanning skills and situation awareness components, in order to see if skills (and subsequently the processing function) prevail
over situation awareness (the storage function) or otherwise.
Figure~\ref{fig_MemGameScore} displays the score at the end of the
games for all agents (clue memory sizes) and all difficulty levels
throughout the tournament. For low difficulty games, the score is
always maximum, with the games being completed regardless of the
size of memory or skill complexity. As the game difficulty
increases, the difference in score between agents with low and
high situation awareness memory becomes significant. The score
drops significantly for agents with high situation awareness
memory, with the drop starting early in the tournament at
difficulty 5. Recalling that high $M_c$ leaves a low amount of
working memory to be used for loading skills ($M_s$), only limited
amount, number and/or complexity, of skills can be used. On the
other hand, results show that the performance is less affected in
the opposite situation, when the situation awareness memory is
low. This suggests that the scarcity of skills can jeopardize
entirely the ability of an agent to complete the game, whereas
severe limitation of situation awareness memory still allows a
certain level of performance. We conclude that the presence of
scanning skills (inherent ability to process) prevails the storage
of situation awareness (ability to store) in the working memory.

\section{Conclusions}\label{04Conclusion}

\noindent We investigated the trade-off between processing and
storage functions of working memory in Sudoku. We used a society
of agents capable of learning from each-other how to effectively
use their existing skills in conjunction with their working memory
in order to solve Sudoku games of various difficulty levels. The
most used skills, and most importantly the effective skills, i.e.
the ones that contribute to acquiring high scores in a Sudoku,
have been established.
The main finding is that, the scanning skills tend to be more
important than the space available for situation awareness.

\section{Acknowledgement}

This project is supported by the Australian Research Council Discovery Grant DP140102590, entitled ``Challenging systems to discover vulnerabilities using computational red teaming". \\
This is a pre-print of an article published in Lecture Notes in Computer Science, vol 8836, Springer. The final authenticated version is available online at: https://doi.org/10.1007/978-3-319-12643-2\_69

\footnotesize
\bibliographystyle{splncs03}

\end{document}